\documentclass[11pt,reqno]{article}
\usepackage{amsmath}
\usepackage{amsfonts}
\usepackage{amssymb}
\usepackage{latexsym}
\usepackage[dvips]{graphicx}
\usepackage{epsf}

\allowdisplaybreaks \numberwithin{equation}{section}

\newcommand{\be}{\begin{equation}}
\newcommand{\ee}{\end{equation}}
\newcommand{\bi}{\begin{itemize}}
\newcommand{\ei}{\end{itemize}}
\newcommand{\bea}{\begin{eqnarray}}
\newcommand{\eea}{\end{eqnarray}}

\let\m=\mu    \let\n=\nu

\let\==\equiv

\newcommand{\Tr}{{\rm Tr}}

\newcommand{\cB}{\mathcal{B}}

\newcommand{\cD}{\mathcal{D}}
\newcommand{\cE}{\mathcal{E}}

\newcommand{\cO}{\mathcal{O}}

\newcommand{\half}{\tfrac{1}{2}}

\newcommand{\lb}{\bar{\lambda}}

\newcommand{\e}{\mathrm{e}}
\newcommand{\p}{\partial}

\begin{document}

\thispagestyle{empty}
\begin{flushright} \small
MZ-TH/11-35
\end{flushright}
\bigskip

\begin{center}
 {\LARGE\bfseries   
   Fractal space-times under the microscope:  \\[1.5ex] 
   A Renormalization Group view on Monte Carlo data    
}
\\[10mm]
Martin Reuter and Frank Saueressig \\[3mm]
{\small\slshape
Institute of Physics, University of Mainz\\
Staudingerweg 7, D-55099 Mainz, Germany \\[1.1ex]
{\upshape\ttfamily reuter@thep.physik.uni-mainz.de} \\
{\upshape\ttfamily saueressig@thep.physik.uni-mainz.de} }\\
\end{center}
\vspace{10mm}

\hrule\bigskip

\centerline{\bfseries Abstract} \medskip
\noindent
The emergence of fractal features in the microscopic structure of space-time is a common theme in many approaches to quantum gravity. In this work we carry out a detailed renormalization group study of the spectral dimension $d_s$ and walk dimension $d_w$ associated with the effective space-times of asymptotically safe Quantum Einstein Gravity (QEG). We discover three scaling regimes where these generalized dimensions are approximately constant for an extended range of length scales: a classical regime where $d_s = d, d_w = 2$, a semi-classical regime where $d_s = 2d/(2+d), d_w = 2+d$, and the UV-fixed point regime where $d_s = d/2, d_w = 4$. On the length scales covered by three-dimensional Monte Carlo simulations, the resulting spectral dimension is shown to be in very good agreement with the data. This  comparison also provides a natural explanation for the apparent puzzle between the short distance behavior of the spectral dimension reported from Causal Dynamical Triangulations (CDT), Euclidean Dynamical Triangulations (EDT), and Asymptotic Safety.
\bigskip
\hrule\bigskip \bigskip \bigskip
 \newpage
\section{Introduction}
Shortly after its introduction, it became clear that the average action approach for gravity \cite{mr} provides strong support for Weinberg's Asymptotic Safety conjecture \cite{wein,wein2}. The key ingredient in this scenario is a non-trivial fixed point of the gravitational renormalization group flow which controls the UV-behavior of the theory and renders Quantum Einstein Gravity (QEG) a non-perturbative renormalizable quantum field theory [1,4-33], see \cite{Reuter:2007rv,oliverbook,livrev,robertorev}  for reviews.  Already at a very early stage of the program various indications pointed in the direction that in this theory space-time should have certain features in common with a fractal. In ref.\ \cite{oliver1} the four-dimensional graviton propagator has been studied in the regime of asymptotically large momenta and it has been found that near the Planck scale a kind of dynamical dimensional reduction occurs. As a consequence of the non-Gaussian fixed point (NGFP) controlling the UV behavior of the theory, the four-dimensional graviton propagator essentially behaves two-dimensional on microscopic scales. 

Subsequently, the "finger prints" of the NGFP on the fabric of the effective QEG space-times have been discussed in \cite{oliver2}, where it was shown that asymptotic safety induces a characteristic self-similarity of space-time on length-scales below the Planck length $\ell_{\rm PL}$. The graviton propagator becomes scale-invariant in this regime \cite{oliver1}, and based on this observation it was argued that, in a cosmological context, the geometry fluctuations it describes can give rise to a scale free spectrum of primordial density perturbations responsible for structure formation \cite{cosmo1,entropy}. Thus the overall picture of the space-time structure in asymptotically safe gravity as it emerged about ten years ago comprises a smooth classical manifold on large distance scales, while on small scales one encounters a low dimensional effective fractal \cite{oliver1,oliver2}. The feature at the heart of these results is the observation that the effective field equations derived from the gravitational average action equip every given smooth space-time manifold with, in principle, infinitely  many different (pseudo) Riemannian structures, one for each coarse graining scale \cite{jan1,jan2}. Thus, very much like in the famous example of the coast line of England \cite{mandel}, the proper length on a QEG space-time depends on the "length of the yardstick" used to measure it.\footnote{Earlier on similar fractal properties had already been found in other quantum gravity theories, in particular near dimension 2 \cite{ninomiya} and in a non-asymptotically safe model \cite{percacci-floreanini-frac}.}

Along a different line of investigations, the Causal Dynamical Triangulation (CDT) approach has been developed and first Monte-Carlo simulations were performed \cite{Ambjorn:2004qm}, see \cite{Ambjorn:2009ts} for a recent review. In this framework one attempts to compute quantum gravity partition functions by numerically constructing the continuum limit of an appropriate statistical mechanics system. From the perspective of the latter, this limit amounts to a second order phase transition. If CDT and its counterpart QEG, formulated in the continuum by means of the average action, belong to the same universality class\footnote{For the time being this is merely a conjecture, of course, albeit a very natural one.} one may expect that the phase transition of the former is described by the non-trivial fixed point underlying the asymptotic safety of the latter.

Remarkably, ref. \cite{janCDT} reported results which indicated that the four-dimensional CDT space-times, too, undergo a dimensional reduction from four to two dimensions as one ``zooms'' in on short distances. In particular it had been demonstrated that the spectral dimension $d_s$ measured in the CDT simulations has the very same limiting behaviors, $4 \rightarrow 2$, as in QEG \cite{oliverfrac}. Therefore it was plausible to assume that both approaches indeed ``see'' the same continuum physics. 

However, this interpretation became problematic when it turned out that the Monte Carlo data correspond to a regime where the cutoff length inherent in the triangulations is still significantly larger than the Planck length. According to the renormalization group (RG) trajectories computed in QEG one would not expect that the asymptotic scaling behavior implied by the fixed point is already realized there \cite{frank1}; on the other hand it is exactly this asymptotic scaling regime to which the QEG prediction of $2$ fractal dimensions pertains \cite{oliver1, oliver2,oliverfrac}. Thus the obvious question is why the CDT simulations detect a significant dimensional reduction despite their appreciable "distance" to the continuum limit.

Recently the situation became even more puzzling. In particular, ref.\ \cite{Benedetti:2009ge} carried out CDT simulations for $d=3$ macroscopic dimensions, which favor a value near $d_s=2$ on the shortest length-scale probed; in this case the QEG prediction for the fixed point region is the value $d_s=3/2$, however \cite{oliverfrac}. Furthermore, the authors of ref.\ \cite{laiho-coumbe} reported simulations within the {\it euclidean} dynamical triangulation (EDT) approach in $d=4$, which favor a drop of the spectral dimension from 4 to about 1.5; this is again in conflict with the QEG expectations if one interprets the latter dimension as the value in the continuum limit.

One of the aims of the present paper is to propose a resolution to these puzzles. In this course we will explicitly compute several types of scale dependent effective dimensions, specifically the spectral dimension $d_s$ and the walk dimension $d_w$ for the effective QEG space-times. We shall see that on length scales slightly {\it larger} than $\ell_{\rm PL}$ there exists a further regime which exhibits the phenomenon of dynamical dimensional reduction. There the spectral dimension is even smaller than near the fixed point, namely $d_s=4/3$ in the case of 4 dimensions classically. Moreover, we shall demonstrate in detail that the (3-dimensional) results reported in \cite{Benedetti:2009ge} are in perfect accord with QEG. In this course, we also verify the supposition \cite{Benedetti:2009ge} that the shortest possible length scale achieved in the simulations is not yet close to the Planck length. Rather the Monte Carlo data probes the transition between the classical and the newly discovered ``semi-classical'' regime.

It is intriguing that Loop Quantum Gravity and spin foam models also show indications for a similar dimensional reduction \cite{modesto2008,modesto-caravelli2009}, with some hints for an intermittent regime where the spectral dimension is smaller than in the deep ultraviolet. In ref.\ \cite{carlip} an argument based upon the strong coupling limit of the Wheeler-DeWitt equation was put forward as a possible explanation of this dimensional reduction. Within non-commutative geometry Connes et al.\ \cite{connes} interpreted the dynamical dimensional reduction to $d_s = 2$ which was observed in QEG in the context of the derivation of the Standard Model from a spectral triple. In fact, from the data provided by a spectral triple, its Dirac operator in particular, one can compute a type of spectral dimension of the resulting non-commutative space which is closely related to the one we are considering here. Also for standard fractals such as Cantor sets, it has been possible to find spectral triples representing them and to compute the corresponding dimensions \cite{spec-trip-frac}. 

Furthermore, a number of model systems (quantum sphere, $\kappa$-Minkowski space, etc.) give rise to a similar reduction as quantum gravity  \cite{Dario-kappa}. Among other developments, these findings also motivated the investigation of physics on {\it prescribed} fractal space-times. In refs. \cite{calcagni-applications,calcagni-etal} a fractional differential calculus \cite{calcagni-reviews} was employed in order to incorporate fractal features, and in  \cite{dunne-phot}  recent exact results on spectral zeta-functions on certain fractals \cite{dunne-complexdim} were used to study the thermodynamics of photons on fractals. In ref.\ \cite{hill} matter quantum field theories were constructed 
and renormalized on a fractal background. This almost universal appearance of fractional properties of space-time and its accessibility in various, a priori different, approaches to Quantum Gravity make the generalized notions of dimensionality discussed in this paper a valuable tool in comparing the physics content of these different formulations.

The remaining parts of this paper are organized as follows. In Sect.\ \ref{Sect.2} we introduce the different notions of ``dimension'', i.e., the spectral, walk, and Hausdorff dimension which will be used to characterize the fractal properties of the effective QEG space-times in Sect.\ \ref{Sect.3}. In Sect.\ \ref{Sect.4} we then analyze how these dimensions change with the RG-scale and identify the scaling regimes where they are approximately constant. These results are then compared with the CDT data obtained in \cite{Benedetti:2009ge} in Sect. \ref{Sect.5}. We conclude with a brief discussion of our findings in Sect.\ \ref{Sect.6}.

\section{Generalized dimensions characterizing fractal space-times}
\label{Sect.2}
Investigating random walks and diffusion processes on fractals, one is led to introduce 
various notions of fractal dimensions, such as the spectral or walk dimension \cite{avra}.
These notions also prove useful when characterizing properties of space-time in quantum gravity, and 
we will review these concepts in the remainder of this section.
\subsection{The spectral dimension}
To start with, consider the diffusion process where a spin-less test particle performs a Brownian random walk on
an ordinary Riemannian manifold with a fixed classical metric $g_{\m\n}(x)$. It is described by the heat-kernel
$K_g(x, x^\prime; T)$ which gives the probability density for a transition of the particle from $x$ to $x^\prime$ during 
the fictitious time $T$. It satisfies the heat equation
\be\label{heateq}
\p_T K_g(x, x^\prime; T) = - \Delta_g K_g(x, x^\prime; T) \, ,
\ee
where $\Delta_g = - D^2$ denotes the Laplace-Beltrami operator. In flat space, this equation is easily solved by
\be\label{heat1}
K_g(x, x^\prime; T) = \int \frac{d^dp}{(2\pi)^d} \, \e^{i p \cdot (x-x^\prime)} \, \e^{-p^2 T} 
\ee
In general, the heat-kernel is a matrix element of the operator $\exp(- T \Delta_g)$. In the random walk picture
its trace per unit volume,
\be
P_g(T) = V^{-1} \int d^dx \sqrt{g(x)} \, K_g(x, x; T) \equiv V^{-1} \, \Tr \, \exp(- T \Delta_g) \, , 
\ee
has the interpretation of an average return probability. Here $V \equiv \int d^dx \sqrt{g(x)}$ denotes the total volume. It is well known that $P_g$ possesses an asymptotic early time expansion (for $T \rightarrow 0$) of the form $P_g(T) = (4 \pi T)^{d/2} \sum_{n=0}^\infty A_n T^n$, with $A_n$ denoting the Seeley-DeWitt coefficients. From this expansion one can motivate the definition of the spectral dimension $d_s$ as the $T$-independent logarithmic derivative
\be \label{DsTlim} 
d_s \equiv \left. - 2 \frac{d \ln P_g(T)}{d \ln T} \right|_{T = 0} \, . 
\ee
On smooth manifolds, where the early time exapnsion of $P_g(T)$ is valid, the spectral dimension agrees with the topological dimension $d$ of the manifold.

Given $P_g(T)$, it is natural to define an, in general $T$-dependent, generalization of the spectral dimension by
\be\label{DsT}
\cD_s(T) \equiv - 2 \frac{d \ln P_g(T)}{d \ln T}\, .
\ee
 According to \eqref{DsTlim}, we recover the true spectral dimension of the space-time by 
considering the shortest possible random walks, i.e., by taking the limit $d_s =  \lim_{T \rightarrow 0} \cD_s(T)$.
Note that in view of a possible comparison with other (discrete) approaches to quantum gravity
 the generalized, scale-dependent version \eqref{DsT} will play a central role later on.

\subsection{The walk dimension}
Regular Brownian motion in flat space has the celebrated property
that the random walker's average square displacement increases linearly
with time: $\langle r^2 \rangle \propto T$. Indeed, performing the integral
\eqref{heat1} we obtain the familiar probability density
\be\label{FSHeat}
K(x, x^\prime; T) = (4 \pi T)^{-d/2} \exp\left(- \frac{\sigma(x, x^\prime)}{2 T} \right)
\ee
with $\sigma(x, x^\prime) = \half |x - x^\prime |^2$ half the squared geodesic distance between the points $x, x^\prime$. 
Using \eqref{FSHeat} yields the expectation value $ \langle r^2 \rangle \equiv \langle x^2 \rangle = \int d^dx \, x^2 \, K(x, 0; T) \propto T$.

Many diffusion processes of physical interest (such as diffusion on fractals) are anomalous in the sense that this linear relationship is generalized to a power law
$\langle r^2 \rangle \propto T^{2/d_w}$ with $d_w \not = 2$. The interpretation of the so-called walk dimension $d_w$ is as follows. The trail left by the random walker is a random object, which is interesting in its own right. It has the properties of a fractal, even in the ``classical'' case when the walk takes place on a regular manifold. The quantity $d_w$ is precisely the fractal dimension of this trail. Diffusion processes are called regular if $d_w = 2$, and anomalous when $d_w \not = 2$.  

\subsection{The Hausdorff dimension}
Finally, we introduce the Hausdorff dimension $d_H$. Instead of working with its mathematical rigorous definition in terms of the Hausdorff measure and all possible covers of the metric space under consideration, the present, simplified definition may suffice for our present purposes. On a smooth set, the scaling law for the volume
$V(r)$ of a $d$-dimensional ball of radius $r$ takes the form 
\be\label{Hdd}
V(r) \propto r^{d_H} \, .
\ee
The Hausdorff dimension is then obtained in the limit of infinitely small radius,
\be
d_H \equiv \lim_{r \rightarrow 0} \frac{\ln V(r)}{\ln r} \, .
\ee
 Contrary to the spectral or walk dimension whose definitions are linked to dynamical diffusion 
 precesses on space-time, there is no dynamics associated with $d_H$.

\section{Fractal dimensions within QEG}
\label{Sect.3}
Upon introducing various concepts for fractal dimensions in the last section,
we now proceed with their evaluation for the 
QEG effective space-times, generalizing the results of ref.\ \cite{oliverfrac}. 
 Our discussion will mostly be based on
the so-called Einstein-Hilbert truncation introduced in the next subsection. As we shall see this restriction
is actually unnecessary in the asymptotic scaling regime, i.e., when the RG-trajectory
is close to the NGFP. In this case we can derive {\it exact} results for the spectral and walk dimension
by exploiting the scale invariance of the theory at the fixed point.

\subsection{Diffusion processes on QEG space-times}
Since in QEG one integrates over all metrics, the central idea is to replace $P_g(T)$ by its expectation value
\be\label{QPgT}
P(T) \equiv \langle P_\gamma(T) \rangle \equiv \int \cD \gamma \cD C \cD \bar{C} \, P_\gamma(T) \, e^{-S_{\rm bare}[\gamma, C, \bar{C}] } \, .
\ee
Here $\gamma_{\m\n}$ denotes the microscopic metric and $S_{\rm bare}$ is the bare action related to the UV fixed 
point, with the gauge-fixing and the pieces containing the ghosts $C$ and $\bar{C}$ included. For the untraced heat-kernel, we define likewise
\be\label{heatexp}
K(x, x^\prime; T) \equiv \langle K_\gamma(x, x^\prime; T) \rangle \, . 
\ee
These expectation values are most conveniently calculated from the effective average action $\Gamma_k$, which equips 
the $d$-dimensional smooth manifolds underlying the QEG effective space-times with a family of metric structures $\left\{ \langle g_{\m\n} \rangle_k, 0 \le k < \infty \right\}$, one for each coarse-graining scale $k$ \cite{oliverfrac,jan1}. These metrics are solutions to the effective field equations implied by $\Gamma_k$. 

To start with, we shall approximate the latter by the Einstein-Hilbert truncation \cite{mr,frank1} 
\be\label{EHtrunc}
\Gamma_k = (16\pi G_k)^{-1} \int d^dx \sqrt{g} \left( - R + 2 \lb_k \right) + \mbox{classical gauge-fixing and ghost terms} \, ,
\ee
which includes a scale-dependent cosmological constant $\lb_k$ and Newtons constant $G_k$.
The corresponding effective field equation reads
\be\label{effeom}
R_{\m\n}(\langle g \rangle_k) = \frac{2}{2-d} \, \lb_k \,  \langle g_{\m\n} \rangle_k \, .
\ee
It has the same form as the classical Einstein equation, with a $k$-dependent cosmological constant $\lb_k$, however. We can easily
find the $k$-dependence of the corresponding solution $\langle g_{\m\n} \rangle_k$ by rewriting \eqref{effeom} as $[\lb_{k_0} / \lb_k] R^\m{}_\n(\langle g \rangle_k) = \frac{2}{2-d} \lb_{k_0} \delta^\m{}_\n $ for some fixed reference scale $k_0$, and exploiting that $R^\m{}_\n(cg) = c^{-1} R^\m{}_\n(g)$ for any constant $c > 0$. This shows that the metric and its inverse scale according to
\be\label{metricscaling}
\langle g_{\m\n}(x) \rangle_k = [\lb_{k_0} / \lb_k] \langle g_{\m\n}(x) \rangle_{k_0} \, , \qquad 
\langle g^{\m\n}(x) \rangle_k = [ \lb_k/ \lb_{k_0}] \langle g^{\m\n}(x) \rangle_{k_0} \, .
\ee
Denoting the Laplace-Beltrami operators corresponding to the metrics $\langle g_{\m\n} \rangle_k$ and $\langle g_{\m\n} \rangle_{k_0}$ by $\Delta(k)$ and $\Delta(k_0)$, respectively, these relations imply
\be\label{Laplacescaling}
\Delta(k) = \left[ \lb_k / \lb_{k_0} \right] \Delta(k_0) \, .
\ee

At this stage, the following remarks are in order. In the asymptotic scaling regime associated with the NGFP the scale-dependence of the couplings is fixed by the fixed point condition:
\be\label{scalingregime}
\lb_k \propto k^2 \, , \qquad G_k \propto k^{2-d} \, .
\ee
This implies in particular
\be\label{UVscaling}
\langle g_{\m\n}(x) \rangle_k \propto k^{-2} \qquad (k \rightarrow \infty) \, .
\ee
This asymptotic relation is actually an {\it exact} consequence of Asymptotic Safety, which solely relies on the scale-independence of the theory at the fixed point.

In general, the relation \eqref{Laplacescaling} will receive corrections from the presence of higher-derivative operators in the effective average action \cite{oliver2,codello,frankmach,BMS}. These contributions organize themselves into a power series controlled by the dimensionless cosmological constant $\lambda_k \equiv \lb^2_k/k^2$ in which the square bracket in \eqref{Laplacescaling} provide the leading term for small values of $\lambda_k$. Thus, for finite scales $k$, the equations \eqref{metricscaling} and \eqref{Laplacescaling} hold true in the Einstein-Hilbert truncation and the absence of matter only. These conditions guarantee that the effective field equation has the simple form \eqref{effeom}, which is necessary for the proportionality $\langle g_{\m\n} \rangle_k \propto \lb_k^{-1}$ expressed in \eqref{scalingregime}. This relation constitutes an essential piece in the derivation of eq.\ \eqref{Laplacescaling}. In the scaling regime of a complete gravity-matter fixed point the conditions above can be relaxed, however, since then for purely dimensional reasons basically, $\langle g_{\m\n} \rangle_k \propto k^{-2}$ in the full theory and any sensible truncation.

This said, we can now evaluate the expectation value \eqref{QPgT} by exploiting the effective field theory properties of the effective average action. Since $\Gamma_k$ defines an effective field theory at the scale $k$ we know that $\langle \cO(\gamma_{\m\n}) \rangle \approx \cO(\langle g_{\m\n} \rangle_k)$ provided the observable $\cO$ involves only momentum scales of the order of $k$. We apply this rule to the RHS of the diffusion equation, $\cO = - \Delta_\gamma K_\gamma(x, x^\prime; T)$. The subtle issue here is the correct identification of $k$. If the diffusion process involves (approximately) only a small interval of scales near $k$ over which $\lb_k$ does not change much, the corresponding heat equation contains the operator $\Delta(k)$ for this specific, fixed value of $k$: $\p_T K(x, x^\prime; T) = - \Delta(k) K(x, x^\prime; T)$. Denoting the eigenvalues of $\Delta(k_0)$ by $\cE_n$ and the corresponding eigenfunctions by $\phi_n$, this equation is solved by
\be\label{heatk}
K(x, x^\prime; T) = \sum_n \phi_n(x) \phi_n(x^\prime) \exp\Big( - F(k^2) \cE_n T \Big) \, . 
\ee 
Here we introduced the convenient notation $F(k^2) \equiv \lb_k/\lb_{k_0}$. Knowing the propagation kernel, we can time-evolve any initial probability distribution $p(x; 0)$ according to 
\be
p(x; T) = \int d^dx^\prime \sqrt{g_0(x^\prime)} \, K(x, x^\prime; T) \, p(x^\prime; 0)
\ee
 with $g_0$ the determinant of $\langle g_{\m\n} \rangle_{k_0}$. If the initial distribution has an eigenfunction expansion of the form $p(x; 0) = \sum_n C_n \phi_n(x)$ we obtain
\be\label{prob2}
p(x; T) = \sum_n C_n \phi_n(x) \exp\Big( - F(k^2) \cE_n T \Big) \, .
\ee

If the $C_n$'s are significantly different from zero only for a single eigenvalue $\cE_N$, we are dealing with a single-scale problem and would identify $k^2 = \cE_N$ as the relevant scale at which the running couplings are to be evaluated. In general the $C_n$'s are different from zero over a wide range of eigenvalues. In this case we face a multiscale problem where different modes $\phi_n$ probe the space-time on different length scales. If $\Delta(k_0)$ corresponds to flat space, say, the eigenfunctions $\phi_n = \phi_p$ are plane waves with momentum $p^\m$, and they resolve structures on a length scale $\ell$ of order $1/|p|$. Hence, in terms of the eigenvalue $\cE_n \equiv \cE_p = p^2$ the resolution is $\ell \approx 1/\sqrt{\cE_n}$. This suggests that when the manifold is probed by a mode with eigenvalue $\cE_n$ it ``sees'' the metric $\langle g_{\m\n} \rangle_k$ for the scale $k = \sqrt{\cE_n}$. Actually, the identification $k = \sqrt{\cE_n}$ is correct also for curved space since, in the construction of $\Gamma_k$, the parameter $k$ is introduced precisely as a cutoff in the spectrum of the covariant Laplacian.

As a consequence, under the spectral sum of \eqref{prob2}, we must use the scale $k^2 = \cE_n$ which depends explicitly on the resolving power of the corresponding mode. Likewise, in eq.\ \eqref{heatk}, $F(k^2)$ is to be interpreted as $F(\cE_n)$:
\be\label{2.19}
\begin{split}
K(x, x^\prime; T) = & \, \sum_n \phi_n(x) \phi_n(x^\prime) \exp\Big(-F(\cE_n) \cE_n T \Big) \\
= & \, \sum_n \phi_n(x^\prime)  \exp\Big(-F\big(\Delta(k_0)\big) \Delta(k_0) T \Big) \phi_n(x^\prime) \, .
\end{split} 
\ee
As in \cite{oliverfrac}, we choose $k_0$ as a macroscopic scale in the classical regime, and we assume that at $k_0$ the cosmological constant is small, so that $\langle g_{\m\n} \rangle_{k_0}$ can be approximated by the flat metric on $\mathbb{R}^d$. The eigenfunctions of $\Delta(k_0)$ are plane waves then and eq.\ \eqref{2.19} becomes
\be\label{Heat:FlatSpace}
K(x, x^\prime; T) = \int \frac{d^dp}{(2 \pi)^d} \, e^{i p \cdot (x-x^\prime)} \, e^{-p^2 F(p^2) T}
\ee
where the scalar products are performed with respect to the flat metric, $\langle g_{\m\n} \rangle_{k_0} = \delta_{\m\n}$. The kernel \eqref{Heat:FlatSpace} satisfies $K(x, x^\prime; 0) = \delta^d(x-x^\prime)$ and, provided that $\lim_{p \rightarrow 0} p^2 F(p^2) = 0$, also $\int d^dx K(x, x^\prime; T) = 1$. 

Taking the normalized trace of \eqref{Heat:FlatSpace} within this ``flat space-approximation'' yields \cite{oliverfrac}
\be\label{2.21}
P(T) = \int \frac{d^dp}{(2 \pi)^d} \, e^{-p^2 F(p^2) T} \, .
\ee
Introducing $z = p^2$, the final result for the average return probability reads
\be\label{2.22}
P(T) = \frac{1}{(4\pi)^{d/2} \Gamma(d/2)} \int_0^\infty dz \, z^{d/2-1} \, \exp\Big(-z F(z) T \Big) \, , 
\ee
where $F(z) \equiv \lb(k^2 = z)/\lb_{k_0}$. 
%
\subsection{The spectral dimension in QEG}
In the classical case, $F(z) = 1$, the relation \eqref{2.22} reproduces the familiar result $P(T) = 1/(4 \pi T)^{d/2}$, whence $\cD_s(T) = d$ independently of $T$.
We shall now discuss the spectral dimension for several other illustrative and important examples.

{\bf (A)} To start with, let us evaluate the average return probability \eqref{2.22} for a simplified RG-trajectory where the scale dependence of the cosmological constant is given by a power law, with the same exponent $\delta$ for all values of $k$:
\be\label{2.30}
\lb_k \propto k^\delta \quad \Longrightarrow \quad F(z) \propto z^{\delta/2} \, . 
\ee
By rescaling the integration variable in \eqref{2.22} we see that in this case
\be\label{2.31}
P(T) = \frac{\rm const}{T^{d/(2+\delta)}} \, . 
\ee
Hence \eqref{DsT} yields the important result
\be\label{powerspect}
\framebox{\; \; $\cD_s(T) = \frac{2d}{2+\delta}$ \bigg. \; \;} \, .
\ee
It happens to be $T$-independent, so that for $T \rightarrow 0$ trivially
\be\label{2.33}
d_s = \frac{2d}{2+\delta} \, . 
\ee

\noindent
{\bf (B)} Next, let us be slightly more general and assume
that the power law \eqref{2.30} is valid only for squared momenta in a 
certain interval, $p^2 \in [z_1, z_2]$, but $\lb_k$ remains unspecified 
otherwise. In this case we can obtain only partial information about $P(T)$,
namely for $T$ in the interval $[z_2^{-1}, z_1^{-1}]$. The reason is that for $T \in [z_2^{-1}, z_1^{-1}]$
the integral in \eqref{2.22} is dominated by momenta for which 
approximately $1/p^2 \approx T$, i.e., $z \in [z_1, z_2]$. This leads us again to the formula
\eqref{powerspect}, which now, however, is valid only for a restricted range of diffusion times $T$; 
in particular the spectral dimension of interest may not be given by extrapolating \eqref{powerspect} to $T \rightarrow 0$.

\noindent
{\bf (C)} Let us consider an arbitrary asymptotically safe RG-trajectory so that its behavior for $k \rightarrow \infty$ is controlled by
the NGFP. In this case the running of the cosmological constant for $k \gtrsim M$, with $M$ a characteristic mass scale of the order of the Planck mass, is given by a quadratic scale-dependence $\lb_k = \lambda_* k^2$, independently of $d$. This corresponds to a power law with $\delta = 2$, which entails in the {\bf NGFP regime}, i.e., for $T \lesssim 1/M^2$,
\be\label{UVspec}
\cD_s(T) = \frac{d}{2} \qquad \quad \Big( \mbox{NGFP regime} \Big) \, . 
\ee 
This dimension, again, is locally $T$-independent. It coincides with the $T \rightarrow 0$ limit:
\be
d_s = \frac{d}{2} \, .
\ee
This is the result first derived in ref.\ \cite{oliverfrac}. As it was explained there, it is actually an exact consequence of Asymptotic Safety
which relies solely on the existence of the NGFP and does not depend on the Einstein-Hilbert truncation.

\noindent
{\bf (D)} Returning to the Einstein-Hilbert truncation, let us consider the piece of the Type IIIa RG-trajectory depicted in Fig.\ \ref{Fig.epl} which lies inside the linear regime of the Gaussian fixed point. Newton's constant is approximately $k$-independent there and the cosmological constant evolves according to
\be\label{2.36}
\lb_k = \lb_0 + \nu G_0 k^d.
\ee
Here $\nu = (4 \pi)^{1-d/2} (d-3) \Phi^1_{d/2}(0)$ is a scheme-dependent constant \cite{mr,frank1}. When
$k$ is not too small, so that $\lb_0$ can be neglected relative to $\nu G_0 k^d$, we are in what we shall call the ``$k^d$ regime''; it is characterized by a pure power law 
$\lb_k \approx k^\delta$ with $\delta = d$. The physics behind this scale dependence is simple and well-known: It represents exactly the vacuum energy density obtained by
summing up the zero point energies of all field modes integrated out. For $T$ in the range of scales pertaining to the $k^d$ regime we find
\be
\cD_s(T) = \frac{2d}{2+d} \qquad (k^d \; \mbox{regime}) \, . 
\ee
Note that for every $d > 2$ the spectral dimension in the $k^d$ regime is even {\it smaller} than in the NGFP regime
\be
\frac{\cD_s(\mbox{NGFP regime})}{\cD_s(k^d \; \mbox{regime})} = 1 + (d-2)/4 \, . 
\ee
%
\subsection{The walk dimension in QEG}
In order to determine the walk dimension for the diffusion on the effective QEG space-times
we return to eq.\ \eqref{Heat:FlatSpace} for the untraced heat-kernel. We restrict ourselves to a regime with 
a power law running of $\lb_k$, whence $F(p^2) = (Lp)^\delta$ with some constant length-scale $L$.

Introducing $q_\m \equiv p_\m T^{1/(2+\delta)}$ and $\xi_\m \equiv (x_\m - x_\m^\prime) / T^{1/(2+\delta)}$ we can rewrite \eqref{Heat:FlatSpace}
in the form
\be\label{2.45}
K(x, x^\prime; T) = \frac{1}{T^{d/(2+\delta)}} \, \Phi\left( \frac{|x-x^\prime|}{T^{1/(2+\delta)}} \right)
\ee
with the function
\be
\Phi(|\xi|) \equiv \int \frac{d^dq}{(2\pi)^d} \, e^{i q \cdot \xi} \, e^{- L^\delta q^{2+\delta}} \, .
\ee
For $\delta = 0$, this obviously reproduces \eqref{FSHeat}. From the argument of $\Phi$ in \eqref{2.45} we infer that $r = |x-x^\prime|$ scales as $T^{1/(2+\delta)}$ so that the walk
dimension can be read off as\footnote{Cf.\ eq.\ (5.18) in ref.\ \cite{avra}.}
\be\label{walkspect}
\framebox{\; \; $\cD_w(T) = 2+\delta$ \bigg. \; \;} \, . 
\ee
In analogy with the spectral dimension, we use the notation $\cD_w(T)$ rather than $d_w$ to indicate that it might refer to an approximate scaling law which is valid for a finite range of scales only.

For $\delta = 0, 2$, and $d$ we find in particular, for any topological dimension $d$, 
\be\label{2.48}
\cD_w = \left\{ 
\begin{array}{cl}
2   & \mbox{classical regime} \\
4   & \mbox{NGFP regime} \\
2+d \, \,  & k^d\mbox{ regime} \\
\end{array}
\right.
\ee
Regimes with all three walk dimensions of \eqref{2.48} can be realized along a single RG-trajectory. Notably, the result for the NGFP regime, $\cD_w = 4$, is exact in the sense, that it does not rely on the Einstein-Hilbert truncation.

\subsection{The Hausdorff dimension in QEG}
The smooth manifold underlying QEG has per se no fractal properties 
whatsoever. In particular, the volume of a $d$-ball $\cB^d$ covering a patch
of the smooth manifold of QEG space-time scales as
\be
V(\cB^d) = \int_{\cB^d} d^dx \sqrt{g_k} \propto (r_k)^d \, . 
\ee 
Thus, by comparing to eq.\ \eqref{Hdd}, we read off that the 
Hausdorff dimension is strictly equal
to the topological one:
\be\label{hdspect}
\framebox{\; \; $d_H = d$ \bigg. \; \;} \, . 
\ee
We emphasize that the effective QEG space-times should {\it not} be visualized as a kind of 
sponge. Their fractal-like properties have no simple geometric interpretation; they are not due
to a ``removing'' of space-time points. Rather they are of an entirely {\it dynamical} nature,
reflecting certain properties of the {\it quantum states} the system ``space-time metric'' can be in.

For standard fractals the quantities $d_s$, $d_w$, and $d_H$ are not independent but are related by 
\cite{orbach}
\be\label{fracrel}
\frac{d_s}{2} = \frac{d_H}{d_w} \, .
\ee
By combining eqs.\ \eqref{powerspect}, \eqref{walkspect}, and \eqref{hdspect} we see that the same
relation holds true for the effective QEG space-times, at least within the Einstein-Hilbert approximation
and when the underlying RG-trajectory is in a regime with power-law scaling of $\lb_k$. For every value
of the exponent $\delta$ we have
\be\label{rel}
\frac{\cD_s(T)}{2} = \frac{d_H}{\cD_w(T)} \, . 
\ee

The results $d_H = d$, $\cD_w = 2 + \delta$ imply that, as soon as
$\delta > d-2$, we have $\cD_w > d_H$ and the random walk is {\it recurrent} then \cite{avra}.
Classically ($\delta = 0$) this condition is met only in low dimensions $d < 2$, but 
in the case of the QEG space-times it is always satisfied in the $k^d$ regime $(\delta = d)$, for example.
So also from this perspective the QEG space-times, due to the specific quantum gravitational dynamics
to which they owe their existence, appear to have a dimensionality smaller than their topological one.

It is particularly intriguing that, in the NGFP regime, $\cD_w = 4$ independently of $d$. Hence the walk
is recurrent $(\cD_w > d_H)$ for $d < 4$, non-recurrent for $d > 4$, and the marginal case $\cD_w = d_H$ is realized
if and only if $d=4$, making $d=4$ a distinguished value. 
Notably, there is another feature of the QEG space-times which singles out $d=4$: It is the only
dimensionality for which $\cD_s$(NGFP regime)$=d/2$ coincides with the effective dimension
$d_{\rm eff} = d + \eta_* = 2$ derived from the graviton propagator \cite{oliver1,oliverfrac}.

The relation \eqref{rel} also has an important implication for a possible 
relation between the QEG effective space-times and those of the CDT approach.
The latter have a non-classical
Hausdorff dimension $d_H \not = d$ on microscopic scales, while $d_H = d$ in QEG.
Hence, by \eqref{rel}, we cannot expect that both $\cD_s$ and $\cD_w$ agree between
CDT and QEG. If it should turn out that actually $\cD_s^{\rm CDT} = \cD_s^{\rm QEG}$
in some non-classical regime, then $\cD_w^{\rm CDT}$ and $\cD_w^{\rm QEG}$ are necessarily
different there.

\section{The RG-flow of $\cD_s$ and $\cD_w$}
\label{Sect.4}
%
We now proceed by discussing the scale-dependence of the spectral and walk dimension.
For this purpose, we consider an arbitrary RG-trajectory $k \mapsto (g_k, \lambda_k)$, where $g_k \equiv G_k k^{d-2}$ and $\lambda_k \equiv \lb_k k^{-2}$ are the dimensionless counterparts
of Newton's constant and the cosmological constant, respectively. Along such a RG-trajectory there might be isolated intervals of $k$-values where the cosmological constant 
evolves according to a power law, $\lb_k \propto k^\delta$, for some constant exponents $\delta$ which are not necessarily the same on different such
intervals. If the intervals are sufficiently long, it is meaningful to ascribe a spectral and walk dimension to them since $\delta = {\rm const}$  implies $k$-independent values 
$\cD_s = 2d/(2+\delta)$ and $\cD_w = 2 + \delta$.

In between the intervals of approximately constant $\cD_s$ and $\cD_w$, where the $k$-dependence of $\lb_k$ is not a power law, 
the notion of a spectral or walk dimension might not be meaningful. The concept of a {\it scale-dependent} dimension
$\cD_s$ or $\cD_w$ is to some extent arbitrary with respect to the way it interpolates between the ``plateaus'' on which $\delta = {\rm const}$  
for some extended period of RG time. While RG methods allow the computation of the $\cD_s$ and $\cD_w$ values on the various plateaus,
it is a matter of convention how to combine them into continuous functions $k \mapsto \cD_s(k), \cD_w(k)$ which interpolate between
the respective values.

\subsection{The exponent $\delta$ as a function on theory space}
In this subsection, we describe a special proposal for a $k$-dependent $\cD_s(k)$ and $\cD_w(k)$ which is motivated by technical simplicity and the general insights
it allows. We retain eqs.\ \eqref{powerspect} and \eqref{walkspect}, but promote $\delta \rightarrow \delta(k)$ to a $k$-dependent quantity 
\be\label{defscale}
\delta(k) \equiv k \p_k \ln(\lb_k) \, .
\ee
When $\lb_k$ satisfies a power law, $\lb_k \propto k^\delta$ this relation reduces to the case of constant $\delta$. If not, $\delta$ has its own scale dependence, but no direct physical interpretation should be attributed to it. The particular definition \eqref{defscale} has the special property that it actually can be evaluated without first solving for the RG-trajectory. The function
$\delta(k)$ can be seen as arising from a certain scalar function on theory space, $\delta = \delta(g, \lambda)$, whose $k$-dependence results from inserting an RG-trajectory: $\delta(k) \equiv \delta(g_k, \lambda_k)$. In fact, \eqref{defscale} implies
$\delta(k) = k \p_k \ln(k^2 \lambda_k) = 2 + \lambda_k^{-1} k \p_k \lambda_k$
so that
$\delta(k) = 2 + \lambda^{-1}_k \beta_\lambda(g_k, \lambda_k)$ upon using the RG-equation $k \p_k \lambda_k = \beta_\lambda(g, \lambda)$. Thus when we consider
$\delta$ as a function on theory space, coordinatized by $g$ and $\lambda$, it reads
\be\label{deltatheoryspace}
\delta(g, \lambda) = 2 + \frac{1}{\lambda} \, \beta_\lambda(g, \lambda) \, . 
\ee
Substituting this relation into
 \eqref{powerspect} and \eqref{walkspect}, the spectral and the walk dimensions become functions
on the $g$-$\lambda$-plane
\be\label{dstheo}
\cD_s(g, \lambda) = \frac{2d}{4 + \lambda^{-1} \beta_\lambda(g, \lambda)} \, , 
\ee
and
\be\label{dwtheo}
\cD_w(g, \lambda) = 4 + \lambda^{-1} \beta_\lambda(g, \lambda) \, .
\ee

To evaluate these expressions further, we use the $\beta$-functions derived in \cite{mr}:
\be\label{betafcts}
\begin{split}
 \beta_\lambda(g, \lambda) = & \, ( \eta_N - 2) \lambda
+ \half \left(4\pi\right)^{1-d/2} g  \\
& \times \left[ 2d(d+1) \Phi^1_{d/2}(- 2 \lambda) - 8d \Phi^1_{d/2}(0) - d (d+1) \eta_N \tilde{\Phi}^1_{d/2}(-2 \lambda) \right] \, , \\
\beta_g(g, \lambda) = & \, (d-2+\eta_N) g \, . 
\end{split}
\ee
Here the anomalous dimension of Newton's constant $\eta_N$ is given by
\be
\eta_N(g, \lambda) = \frac{g B_1(\lambda)}{1 - g B_2(\lambda)}
\ee
with the following functions of the dimensionless cosmological constant:
\be
\begin{split}
B_1(\lambda) \equiv & \,  \tfrac{1}{3} \left( 4 \pi \right)^{1-d/2}  \Big[
d(d+1)\Phi^1_{d/2-1}(-2\lambda) - 6 d (d-1) \Phi^2_{d/2}(-2\lambda)    \\
& \qquad \qquad \qquad - 4 d \Phi^1_{d/2-1}(0) - 24 \Phi^2_{d/2}(0) \Big] \, ,\\
B_2(\lambda) \equiv & \, - \tfrac{1}{6} (4 \pi)^{1-d/2} \left[d(d+1) \tilde{\Phi}^1_{d/2-1}(-2\lambda) - 6 d (d-1) \tilde{\Phi}^2_{d/2}(-2 \lambda) \right] \, . 
\end{split}
\ee
For practical computations we use the threshold functions resulting from the optimized cutoff
\be\label{opt}
\Phi^p_n(w) = \frac{1}{\Gamma(n+1)} \frac{1}{(1+w)^p} \, , \qquad \tilde{\Phi}^p_n(w) = \frac{1}{\Gamma(n+2)} \frac{1}{(1+w)^p} \, . 
\ee
\begin{figure}[t]
  \centering
    \includegraphics[width=0.77\textwidth]{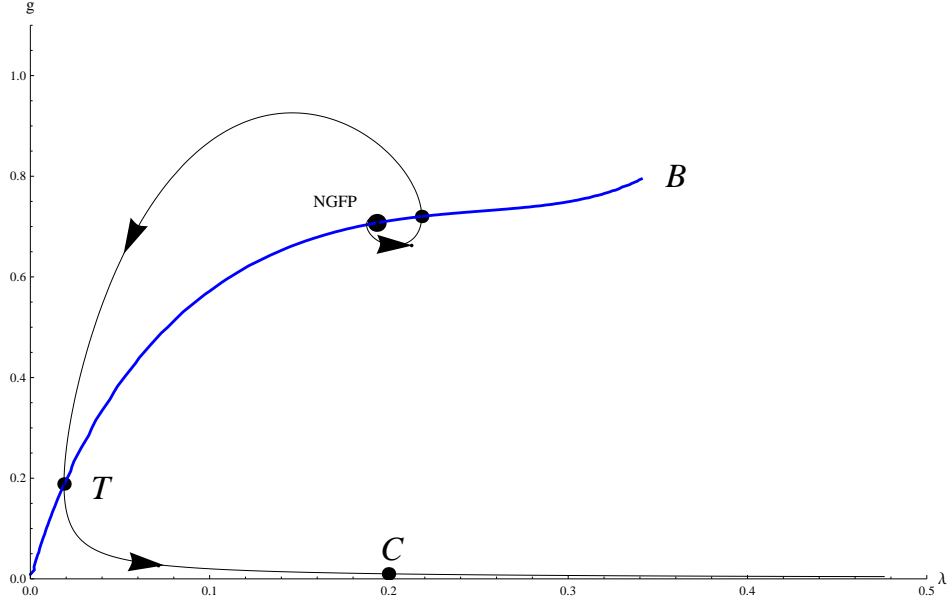}
 \caption{The $g$-$\lambda-$theory space with the line of turning points, $\cB$, and a typical trajectory of Type IIIa. The arrows point
in the direction of decreasing $k$. The big black dot indicates the NGFP while the smaller dots represent points at which the RG-trajectory switches
from increasing to decreasing $\lambda$ or vice versa. The point $T$ is the lowest turning point, and $C$ is a typical point within the classical regime. For $\lambda \gtrsim 0.4$, the RG-flow leaves the classical regime and is no longer reliably captured by the Einstein-Hilbert truncation.}
  \label{Fig.epl}
\end{figure}

As we discussed already, the scaling regime of a NGFP has the exponent $\delta = 2$. From eq.\ \eqref{deltatheoryspace} we learn that this value is realized at all points $(g, \lambda)$ where $\beta_\lambda =0$. The second condition for the NGFP, $\beta_g = 0$, is not required here, so that we have $\delta = 2$ along the entire line in theory space:
\be\label{2.75}
\cB = \Big\{ \, (g, \lambda) \, \Big| \, \beta_\lambda(g, \lambda) = 0 \, \Big\} \, . 
\ee
For $d=4$ the curve $\cB$ is shown as the bold blue line in Fig.\ \ref{Fig.epl}. Both the Gaussian fixed point (GFP) $(g, \lambda) = (0, 0)$ and the NGFP, $(g, \lambda) = (g_*, \lambda_*)$, are located on this curve.\footnote{At the GFP $\cD_s = d$, $\cD_w = 2$, however, since at this point both $\lambda = 0, \beta_\lambda = 0$ so that $\lambda^{-1} \beta_\lambda |_{\rm GFP} = - 2$.} Furthermore, the turning points $T$ of all Type IIIa trajectories are also situated on $\cB$, and the same holds for all the higher order turning points which occur when the trajectory spirals around the NGFP. This observation leads us to an important conclusion: The values $\delta = 2 \Longleftrightarrow  \cD_s = d/2, \cD_w = 4$ which (without involving any truncation) are found in the NGFP regime, actually also apply to all points $(g, \lambda) \in \cB$, provided the Einstein-Hilbert truncation is reliable and no matter is included. 
\begin{figure}[t]
  \centering
    \includegraphics[width=0.75\textwidth]{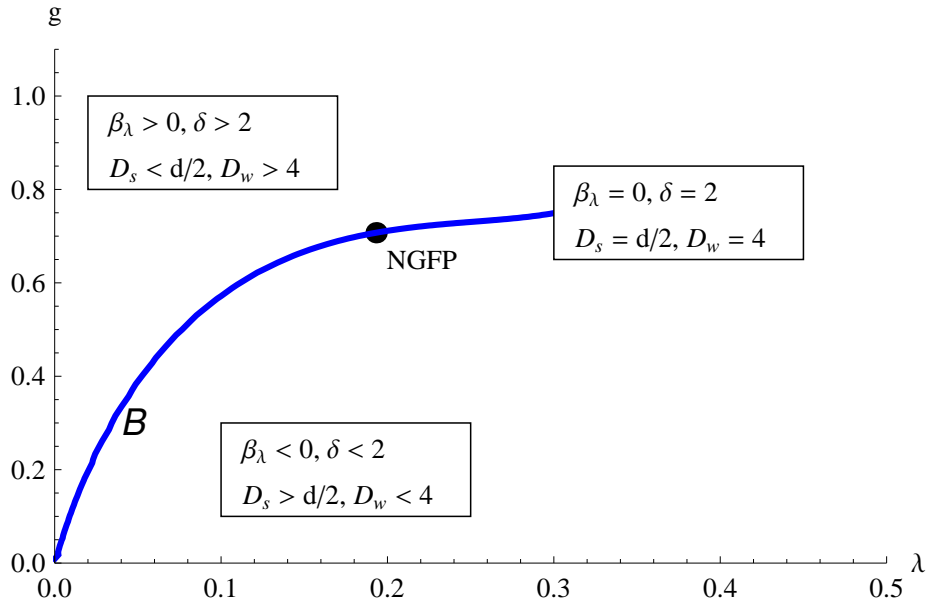}
 \caption{The line of turning points $\cB$ divides the $g$-$\lambda$-plane in two parts on which
$\cD_s$ and $\cD_w$ are everywhere either larger or smaller than at the NGFP.}
  \label{Fig.epl2}
\end{figure}
%

\subsection{The spectral and walk dimensions along a RG-trajectory}
We proceed by investigating how the spectral and walk dimension of the effective QEG space-times changes along a given RG-trajectory.
As discussed above, our interest is in scaling regimes where $\cD_s$ and $\cD_w$ remain (approximately) constant for a long interval of $k$-values.
For the remainder of this subsection, we will restrict ourselves to the case $d=4$ for concreteness. 

We start by numerically solving the coupled differential equations
\be\label{EHflow}
k \p_k g(k) = \beta_g( g(k), \lambda(k)) \, , \qquad k \p_k \lambda(k) = \beta_\lambda(g(k), \lambda(k)) \, ,
\ee
with the $\beta$-functions \eqref{betafcts} for a series of initial conditions keeping $\lambda_{\rm init} = \lambda(k_0) = 0.2$ fixed and successively lowering
$g_{\rm init} = g(k_0)$. The result is a family of RG-trajectories where the classical regime becomes more and more pronounced. Subsequently, these solutions are substituted into \eqref{dstheo} and \eqref{dwtheo}, which give $\cD_s(t; g_{\rm init}, \lambda_{\rm init})$ and $\cD_w(t; g_{\rm init}, \lambda_{\rm init})$ in dependence of the RG-time $t \equiv \ln(k)$ and the RG-trajectory. One can verify explicitly, that substituting the RG-trajectory into the return probability \eqref{2.22} and computing the spectral dimension from \eqref{DsTlim} by carrying out the resulting integrals numerically gives rise to the same picture.

\begin{figure}[t]
  \centering
    \includegraphics[width=0.45\textwidth]{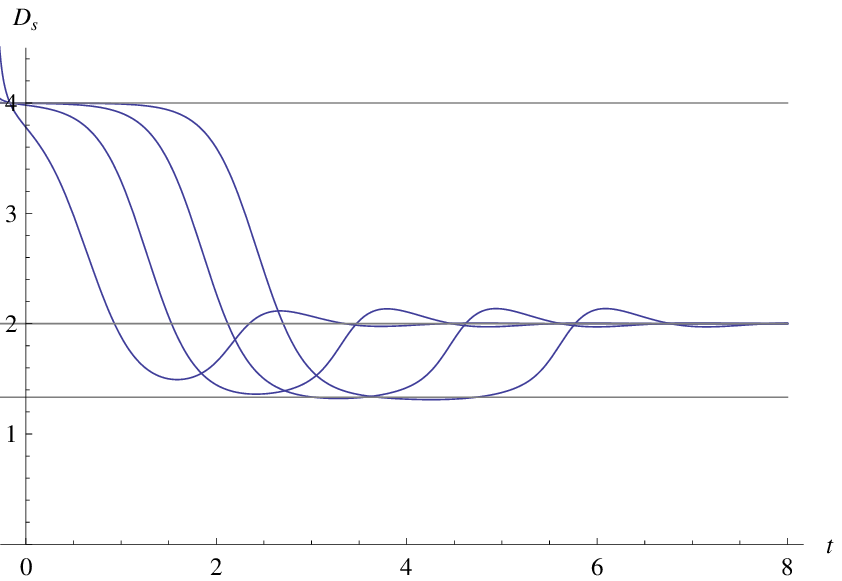} \, \, \, 
    \includegraphics[width=0.45\textwidth]{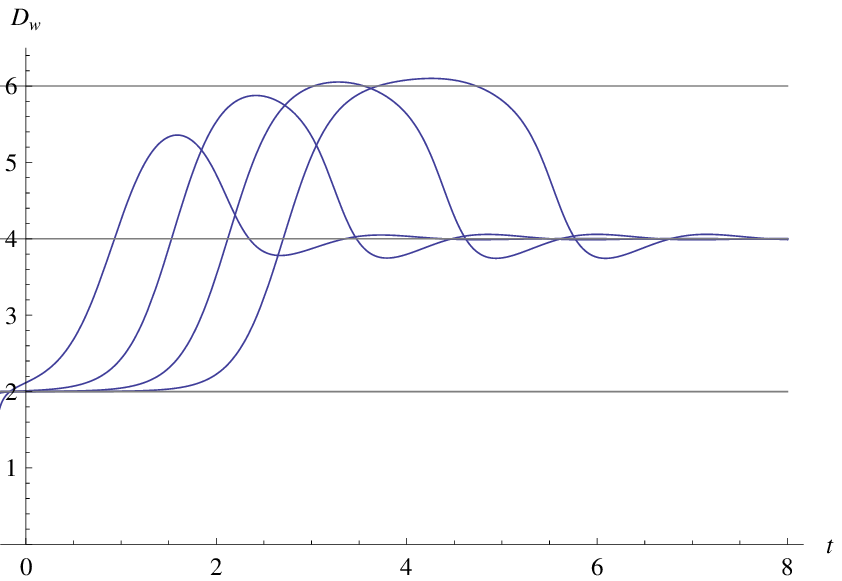} \\[1.2ex]
    \includegraphics[width=0.45\textwidth]{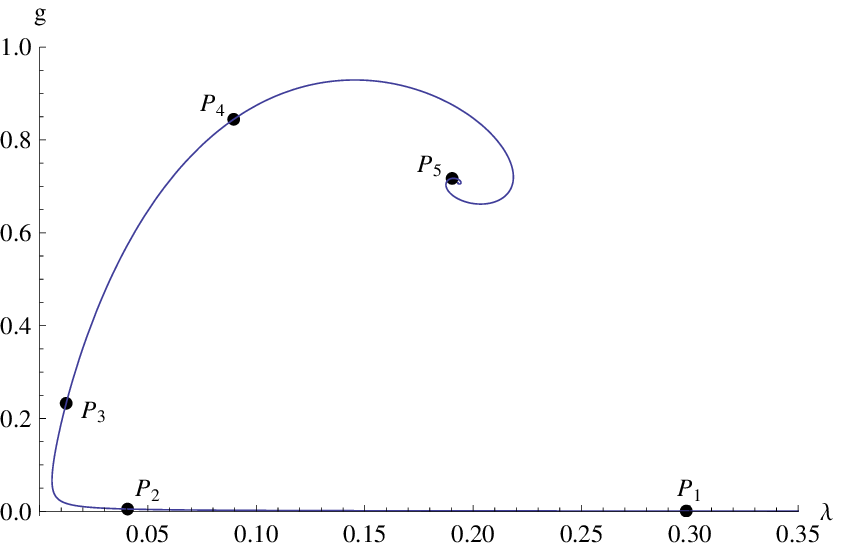}
 \caption{The $t \equiv \ln(k)$-dependent spectral dimension (upper left) and walk dimension (upper right) along illustrative solutions of the RG-equations \eqref{EHflow} in $d=4$. The trajectories develop three plateaus: the classical plateau with $\cD_s =4, \cD_w = 2$, the semi-classical plateau where $\cD_s = 4/3, \cD_w = 6$ and the NGFP plateau with $\cD_s = 2, \cD_w = 4$. These plateau values are indicated by the gray horizontal lines and connected by crossover parts. The lower figure shows the location of these plateaus on the RG-trajectory: the classical, $k^4$, and NGFP regime appear between the points $P_1$ and $P_2$, $P_3$ and $P_4$, and above $P_5$, respectively.}
  \label{Fig.spec}
\end{figure}

Fig.\ \ref{Fig.spec} then shows the resulting spectral dimension, the walk dimension, and the localization of the plateau-regimes on the RG-trajectory in the top-left, top-right and lower diagram, respectively. In the top diagrams, $g_{\rm init}$ decreases by one order of magnitude for each shown trajectory, starting with the highest value to the very left.
As a central result, Fig.\ \ref{Fig.spec} establishes that the RG-flow gives rise to {\it three} plateaus where $\cD_s(t)$ and $\cD_w(t)$ are approximately constant: \\
{\bf (i)} For small values $k$, below $t \simeq 1.8$, say, one finds a {\it classical plateau} where $\cD_s = 4, \cD_w = 2$ for a long range of $k$-values. Here $\delta = 0$, indicating that the cosmological constant is indeed constant. \\
{\bf (ii)} Following the RG-flow towards the UV (larger values of $t$) one next encounters the {\it semi-classical plateau} where $\cD_s = 4/3, \cD_w = 6$. In this case $\delta(k) = 4$ so that $\lb_k \propto k^4$ on the corresponding part of the RG-trajectory. \\
{\bf (iii)} Finally, the {\it NGFP plateau} is characterized by $\cD_s = 2, \cD_w = 4$, which results from the scale-dependence of the cosmological constant at the NGFP $\lb_k \propto k^2 \Longleftrightarrow  \delta = 2$. \\

At this stage, it is worthwhile to see which parts of a typical RG-trajectory realize the scaling relations underlying the plateau-values of $\cD_s$ and $\cD_w$. This is depicted in the lower diagram of Fig.\ \ref{Fig.spec} where we singled out the third solution with $g_{\rm init} = 10^{-3}$ for illustrative purposes. In this case the classical plateau is bounded by the points $P_1$ and $P_2$ and appears well below the turning point $T$, while the semi-classical plateau is situated between the points $P_3$ and $P_4$ well above the turning point. The NGFP plateau is realized by the piece of the RG-trajectory between $P_5$ and the NGFP. The turning point $T$ is not situated in any scaling region but appears along the crossover from the classical to the semi-classical regime of the QEG space-times. For $t < 0$, the spectral dimension (walk dimension) increases (decreases) rapidly. In this region, the underlying RG-trajectory is evaluated outside the classical regime at points $\lambda \gtrsim 0.35$. In this region of the theory space, the Einstein-Hilbert truncation is no longer trustworthy, so that this rapid increase of $\cD_s$ is most likely an artefact, arising from the use of an insufficient truncation.

Notably, the plateaus observed above become more and more extended the closer the trajectories turning point $T$ gets to the GFP, i.e., the smaller the IR value of the cosmological constant. The first RG-trajectory with the largest value $g_{\rm init} = 0.1$ does not even develop a classical and semi-classical plateau, so that a certain level of fine-tuning of the initial conditions is required in order to make these structures visible. Interestingly enough, when one matches the observed data against the RG-trajectories of the Einstein-Hilbert truncation \cite{h3,entropy} one finds that the 
``RG-trajectory realized by Nature'' displays a very extreme fine-tuning of this sort. The coordinates of the turning point are approximately $g_{T} \approx \lambda_{T} \approx 10^{-60}$ and it is passed at the scale $k_{T} \approx 10^{-30} m_{\rm Pl} \approx 10^{-2} {\rm eV} \approx (10^{-2} {\rm mm})^{-1}$, so that there will be very pronounced plateau structures in this case.

\section{Matching the spectral dimensions of QEG and CDT}
\label{Sect.5}
The key advantage of the spectral dimension $\cD_s(T)$ is that it may 
be defined and computed within various a priori unrelated approaches to
quantum gravity. In particular, it is easily accessible in Monte Carlo simulations
of the Causal Dynamical Triangulations (CDT) approach in $d=4$ \cite{janCDT} and $d=3$ \cite{Benedetti:2009ge} as well as
in Euclidean Dynamical Triangulations (EDT) \cite{laiho-coumbe}. This feature allows a direct comparison between
 $\cD_s^{\rm CDT}(T)$ and $\cD_s^{\rm EDT}(T)$ obtained within the discrete approaches and $\cD_s^{\rm QEG}(T)$ capturing the fractal properties of the QEG effective space-times.
In this section we will carry out this comparison for $d=3$. In particular we shall determine the specific
RG-trajectory of QEG which, we believe, underlies the numerical data obtained in \cite{Benedetti:2009ge}.
In principle, it is straightforward to do the same comparison
in $d=4$. This, however, will require access to the detailed Monte Carlo data produced by the four-dimensional CDT or EDT simulations.\footnote{We thank D.\ Benedetti and J.\ Henson for sharing 
the Monte Carlo data underlying their work \cite{Benedetti:2009ge} with us.}

\begin{figure}[t]
  \centering
    \includegraphics[width=0.42\textwidth]{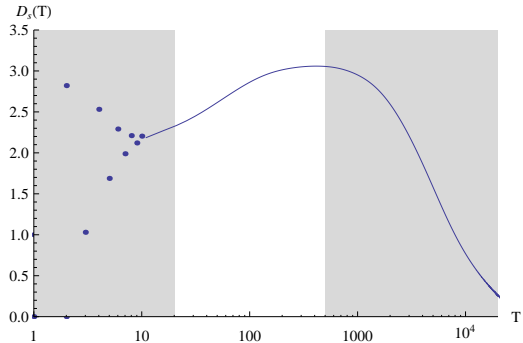}
\caption{Spectral dimension $\cD_s^{\rm CDT}(T)$ determined from random walks on a CDT space-time built from $N = 200$k simplices \cite{Benedetti:2009ge}.}
  \label{Fig.CDT}
\end{figure}

Let us start by looking into the typical features of the spectral dimension $\cD_s^{\rm CDT}(T)$ obtained from the simulations. A prototypical data set showing 
$\cD_s^{\rm CDT}(T)$ as function of the length of the random walk $T$ is given in Fig.\ \ref{Fig.CDT}.
The resulting curve is conveniently split into three regimes: \\
{\bf (i)} For $T \le 20$, corresponding to the left gray region in Fig.\ \ref{Fig.CDT}, $\cD_s^{\rm CDT}(T)$ undergoes rapid oscillations. These originate from the discrete structure of the triangulation to which the short random walks are particular sensitive. \\
{\bf (ii)} For long random walks with $T \ge 500$, the data shows an exponential fall-off. This feature is due to the compact nature of the triangulation, which implies that for long random walks $\cD_s^{\rm CDT}(T)$ is governed by the lowest eigenvalue of the Laplacian on the compact space. This regime is marked by the right gray region in Fig.\ \ref{Fig.CDT}. \\
{\bf (iii)} Between these two regimes, $\cD_s^{\rm CDT}(T)$ is affected neither by the discreteness nor the compactness of the triangulation. Since for $\cD_s^{\rm QEG}(T)$, determined by the flat-space approximation \eqref{2.21}, we do not expect any of these effects to appear, we use this middle region to compare the $T$-dependent spectral dimensions arising from the two, a priori different, approaches. 

This comparison is then carried out as follows: \\
{\bf (i)} First, we numerically construct a RG-trajectory $g_k(g_0, \lambda_0), \lambda_k(g_0, \lambda_0)$ depending on the initial conditions $g_0, \lambda_0$, by solving the flow equations \eqref{EHflow}. \\
{\bf (ii)} Subsequently, we evaluate the resulting spectral dimension $\cD_s^{\rm QEG}(T; g_0, \lambda_0)$ of the corresponding effective QEG space-time.
This is done by first finding the return probability $P(T; g_0, \lambda_0)$, eq.\ \eqref{2.22}, for the RG-trajectory under consideration and then substituting the resulting expression into \eqref{DsT}.  Besides on the length of the random walk, the spectral dimension constructed in this way also depends on the initial conditions of the RG-trajectory. \\ 
{\bf (iii)} Finally, we determine the RG-trajectory underlying the CDT-simulations by 
fitting the parameters $g_0, \lambda_0$ to the Monte Carlo data. The corresponding best-fit values are obtained via an ordinary least-square fit, minimizing the squared Euclidean distance 
\be\label{lsf}
(\Delta \cD_s )^2 \equiv \sum_{T = 20}^{500} \, \left( \cD_s^{\rm QEG}(T; g_0^{\rm fit}, \lambda_0^{\rm fit}) - \cD_s^{\rm CDT}(T) \right)^2 \, , 
\ee
 between the (continuous) function $\cD_s^{\rm QEG}(T; g_0, \lambda_0)$ and the points $\cD_s^{\rm CDT}(T)$. We thereby restrict ourselves to 
the random walks with discrete, integer length $20 \le T \le 500$, which constitute the white part of Fig.\ \ref{Fig.CDT} and correspond to the regime {\bf (iii)} discussed above. 

\begin{table}
\begin{center}
\begin{tabular}{r | c c | c | }
        & \quad \qquad $g_0^{\rm fit}$ \qquad \quad & \quad \qquad $\lambda_0^{\rm fit}$ \quad \qquad & \qquad \quad  \quad $(\Delta \cD_s )^2$ \quad  \qquad \quad \\ \hline
 $70$k  &  $0.7 \times 10^{-5}$ & $ 7.5 \times 10^{-5}$  & $0.680$ \\
 $100$k &  $8.8 \times 10^{-5}$ & $39.5 \times 10^{-5}$  & $0.318$ \\
 $200$k &  $13 \times 10^{-5}$  & $61 \times 10^{-5}$    & $0.257$ \\ \hline
\end{tabular}
\end{center}
\caption{\label{Table.1} Initial conditions $g_0^{\rm fit}, \lambda_0^{\rm fit}$ for the RG-trajectory providing the best fit to the Monte Carlo data \cite{Benedetti:2009ge}. The fit-quality $(\Delta \cD_s )^2$, given by the sum of the squared residues, improves systematically when increasing the number of simplices in the triangulation.}
\end{table}
The resulting best-fit values $g_0^{\rm fit}, \lambda_0^{\rm fit}$ for the triangulations with $N = 70.000$, $N=100.000$, and $N=200.000$ simplices are collected in Table \ref{Table.1}.
Notably, the sum over the squared residuals in the third column of the table  
 improves systematically with an increasing number of simplices. By integrating
 the flow equation for $g(k), \lambda(k)$ for the best-fit initial conditions one furthermore observes that the points $g_0^{\rm fit}, \lambda_0^{\rm fit}$ are actually located 
on {\it different} RG-trajectories. Increasing the size of the simulation $N$ leads to a mild, but systematic increase of the distance between the turning point $T$ and the GFP of the corresponding best-fit trajectories.

\begin{figure}[t]
  \centering
    \includegraphics[width=0.42\textwidth]{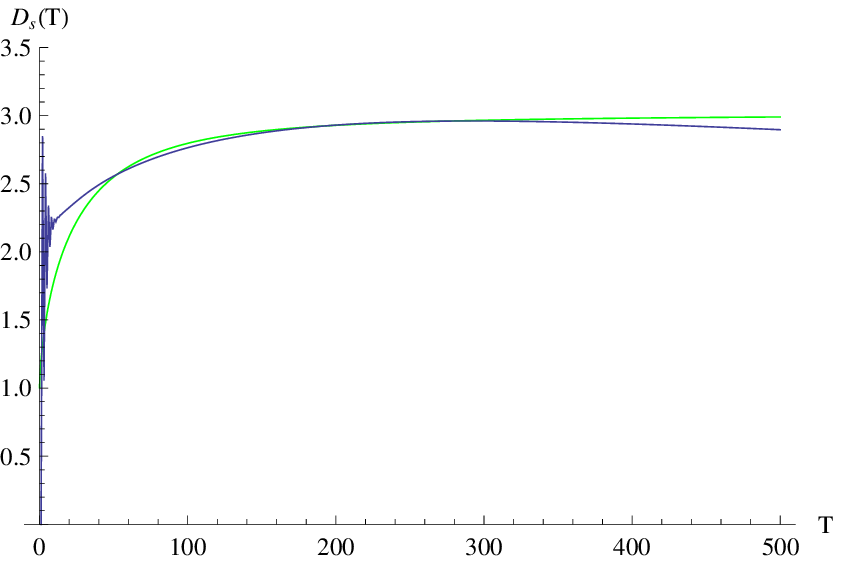} \;\;\;
    \includegraphics[width=0.42\textwidth]{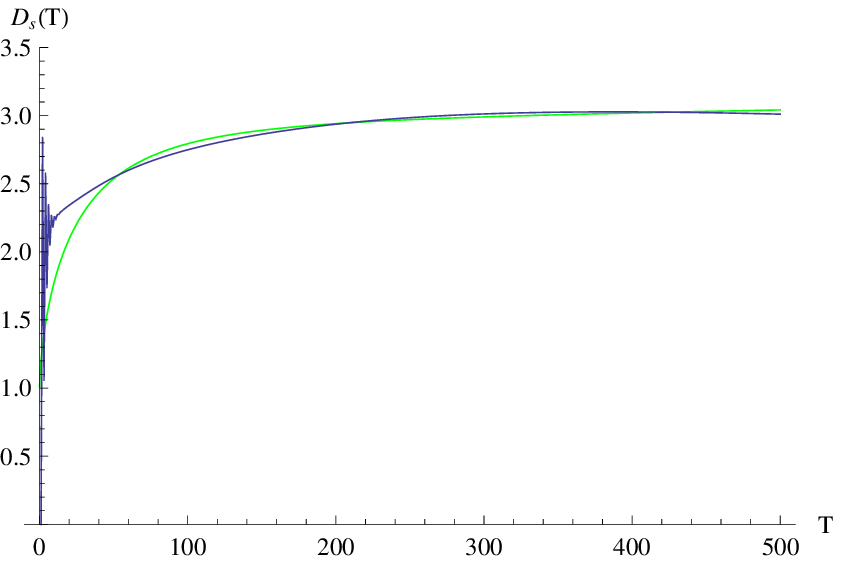} \\
   \includegraphics[width=0.42\textwidth]{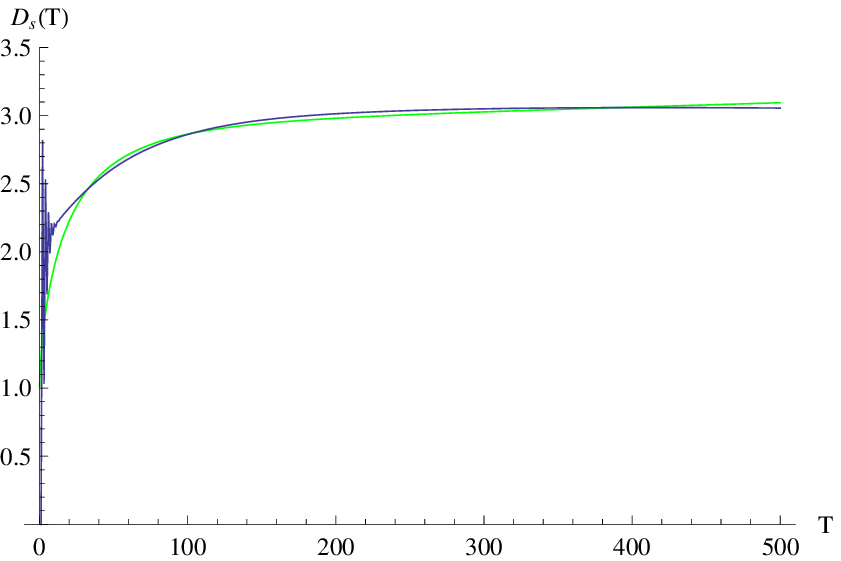} \;\;\;
    \includegraphics[width=0.42\textwidth]{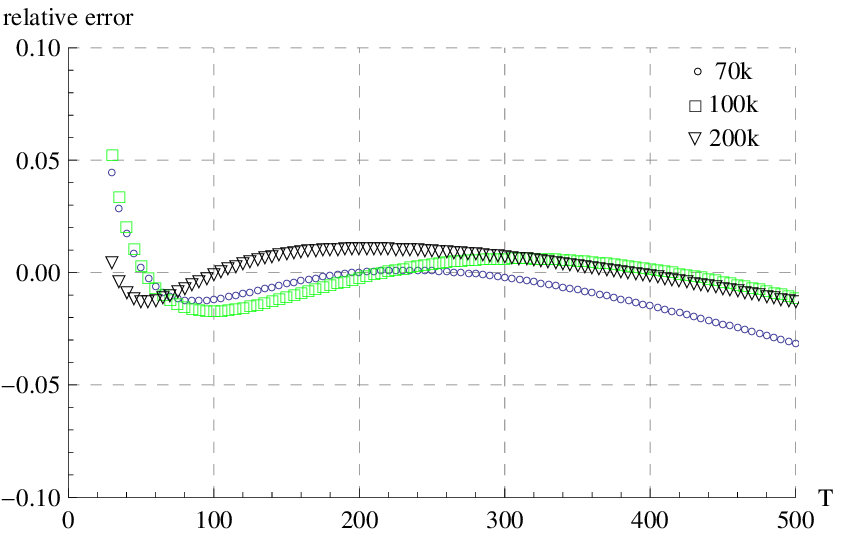}
   \caption{Comparison between the 3-dimensional CDT data-sets $70$k (upper left), $100$k (upper right), and $200$k (lower left) obtained in \cite{Benedetti:2009ge} (blue curves) and the best fit values for $\cD_s^{\rm QEG}(T; g_0^{\rm fit}, \lambda_0^{\rm fit})$ (green curves). The relative errors for the fits to the CDT-datasets with $N = 70.000$ (circles), $N=100.000$ (squares) and $N = 200.000$ (triangles) simplices are shown in the lower right. The residuals grow for very small and very large durations $T$ of the random walk, consistent with discreteness effects at small distances and the compactness of the simulation for large values of $T$, respectively. The quality of the fit improves systematically for triangulations containing more simplices. For the $N=200$k data the relative error is $\approx 1\%$.}
  \label{p.fit1}
\end{figure}

Fig.\ \ref{p.fit1} then shows the direct comparison between the spectral dimensions obtained by the simulations (blue curves) and the best-fit QEG trajectories (green curves)
for $70$k, $100$k and $200$k in the upper left, upper right and lower left panel, respectively. 
 This data is complemented by the relative error 
\be
\epsilon \equiv - \frac{\cD_s^{\rm QEG}(T; g_0^{\rm fit}, \lambda_0^{\rm fit}) - \cD_s^{\rm CDT}(T)}{\cD_s^{\rm QEG}(T; g_0^{\rm fit}, \lambda_0^{\rm fit})}
\ee
for the three fits in the lower right panel. The $70$k data still shows a systematic deviation from the classical value $\cD_s(T) = 3$ for long random walks, which is not present in the QEG results. This mismatch decreases systematically for larger triangulations where the classical regime becomes more and more pronounced. Nevertheless and most remarkably we find that for the $200$k-triangulation that $\epsilon \lesssim 1\%$, throughout. All three sets of residues thereby show a systematic oscillatory structure. These originate from tiny oscillations in the CDT data which are not reproduced by $\cD_s^{\rm QEG}(T)$. Such oscillations commonly appear in systems with discrete symmetries \cite{calcagni-reviews} and are thus likely to be absent in the continuum computation. As a curiosity, we observe that the QEG result matching the most extensive simulation with $N = 200$k ``overshoots'' the classical value $\cD_s(T) = 3$, yielding $\cD_s^{\rm QEG}(T) > 3$ for $T \gtrsim 450$. At this stage, the RG-trajectory is evaluated outside the classical regime in a region of theory space where the Einstein-Hilbert approximation starts to become unreliable. It is then tempting to speculate that larger triangulations may also be sensitive to quantum gravity effects at distances beyond the classical regime.

\begin{figure}[t]
  \centering
    \includegraphics[width=0.72\textwidth]{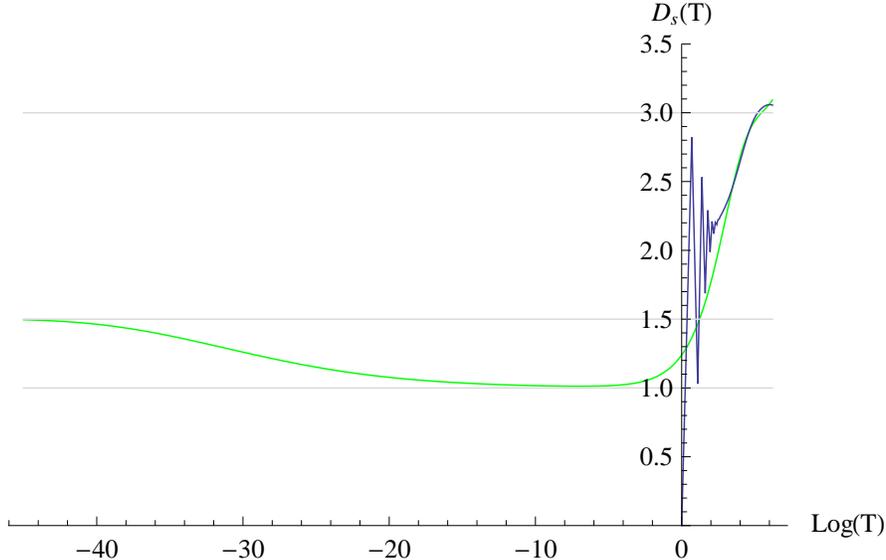}
   \caption{Comparison between the spectral dimensions obtained from the dynamical triangulation with $200$k simplices (blue curve) and the corresponding $\cD_s^{\rm QEG}(T; g_0^{\rm fit}, \lambda_0^{\rm fit})$ predicted by QEG (green curve). In the latter case, the scaling regime corresponding to the NGFP is reached for $\log(T) < -40$, which is well below the distance scales probed by the Monte Carlo simulation.}
  \label{p.fit2}
\end{figure}

We conclude this section by extending $\cD_s^{\rm QEG}(T; g_0^{\rm fit}, \lambda_0^{\rm fit})$ obtained from the $200$k data to the region of very short random walks $T < 20$. The result is depicted in Fig.\ \ref{p.fit2} which displays $\cD_s^{\rm CDT}(T)$ (blue curve) and $\cD_s^{\rm QEG}(T; g_0^{\rm fit}, \lambda_0^{\rm fit})$ (green curve) as a function of $\log(T)$. 
Similarly to the four-dimensional case discussed in Fig.\ \ref{Fig.spec}, the function $\cD_s^{\rm QEG}(T; g_0^{\rm fit}, \lambda_0^{\rm fit})$ obtained for $d=3$ develops three plateaus where the spectral dimension is approximately constant over a long $T$-interval. For successively decreasing duration of the random walks, these plateaus correspond to the classical regime $\cD_s^{\rm QEG}(T) = 3$, the semi-classical regime where $\cD_s^{\rm QEG}(T) \approx 1$ and the NGFP regime where $\cD_s^{\rm QEG}(T) = 3/2$. The figure illustrates that $\cD_s^{\rm CDT}(T)$ probes the classical regime and part of the first crossover towards the semi-classical regime only. This is in perfect agreement with the assertion \cite{Benedetti:2009ge} that the present simulations do not yet probe structures below the Planck scale.

\section{Discussion and conclusions}
\label{Sect.6}
In this work we analyzed the fractal properties of the effective space-times arising within Quantum Einstein Gravity (QEG) formulated in the continuum by means of 
the gravitational average action. These effects are, to some extent, encapsulated in the spectral, walk, and Hausdorff dimension seen by a fictitious 
diffusion process set up on the effective space-times. Most remarkably, these generalized dimensions are found to 
depend on the length of the random walk, indicating that the effective QEG space-times possess a multifractal structure. In particular, we established the possibility of a 
``low energy fractality'' which occurs already well below the asymptotic scaling regime governed by the non-Gaussian (UV) fixed point (NGFP) and is thus unrelated to 
 Asymptotic Safety.  

In Sections \ref{Sect.3} and \ref{Sect.4} we studied this multifractal structure within the flat-space Einstein-Hilbert approximation to QEG.
Thereby it turned out that the effective QEG space-times are comparable to standard fractals in the sense that the relation \eqref{fracrel}
is satisfied on all scales. Their Hausdorff dimension is constant and equal to the topological dimension of the (background) space-time. 
Thus the fractal properties do not originate from the QEG space-times ``loosing points'' at short distances but rather represent a genuine dynamical effect of quantum field theory.
 In contrast to the Hausdorff dimension the spectral dimension 
and the walk dimension seen by the diffusion process depend on the diffusion time $T$. In Fig.\ \ref{Fig.spec} we identified \emph{three} regimes in which
 these generalized dimensions are constant for a wide range of scales. These are connected by short crossovers. For long random walks, QEG space-times have the same spectral properties as classical flat space,
i.e., the diffusion process is regular, $\cD_w = 2$, and the spectral dimension matches the canonical dimension $\cD_s = d$. 
Moving towards shorter diffusion times one encounters the semi-classical scaling regime where $\cD_w = 2 + d$, $\cD_s = 2d/(2+d)$. 
For infinitesimal random walks $T \rightarrow 0$, the properties of the effective space-times are controlled by the NGFP and we obtain
 $\cD_w = 4$ and $\cD_s = d/2$. On both of the latter plateaus the random walk is recurrent.

While the results concerning the NGFP regime are exact and follow directly from the very existence of the fixed point, the ``low energy'' properties of the function
$\cD_s(T)$ rely on the applicability of the approximate effective field equation \eqref{EHtrunc}. It may be used if the Einstein-Hilbert approximation is
sufficiently precise, and if no matter fields are coupled to gravity whose energy-momentum tensor could possibly dominate over the cosmological constant term.
In fact, our derivation of the running effective dimensions made essential use of the proportionality $\langle g_{\m\n} \rangle_k \propto 1/\lb_k$
which does follow from the vacuum Einstein-equation \eqref{EHtrunc}, but not necessarily from a more complicated field equation
$G^k_{\m\n}(\langle g \rangle_k) = - \lb_k \langle g_{\m\n} \rangle_k - 8 \pi G_k T^k_{\m\n}(\langle g \rangle_k)$ 
where $G^k_{\m\n}$ is a (higher derivative, etc.) generalization of the Einstein tensor, and $T^k_{\m\n}(\langle g \rangle_k)$
the energy-momentum tensor of the matter system in the effective geometry described by $(\langle g_{\m\n} \rangle_k)$. The 
multifractal properties at length-scales below the asymptotic NGFP regime explored in this paper can occur only if the curvature of space-time 
is governed by precisely the scale-dependence of $\lb_k$. If, on the other hand, $\lb_k \langle g_{\m\n} \rangle_k$ is negligible as compared
to a $k$-independent term in $G_k T^k_{\m\n}(\langle g \rangle_k)$ the effective geometry has no significant scale dependence
and hence no fractal features.\footnote{For the sake of the argument we assume that $G_{\m\n}^k \equiv G_{\m\n}$ is the conventional Einstein tensor here.} For this reason 
we expect that in real Nature the onset of the fractal behavior is typically shifted towards considerably higher energy scales than expected from the pure gravity case discussed at the end of Section \ref{Sect.4}. We hope to come back to this point in a future publication.

In Sect.\ \ref{Sect.5} we performed a direct comparison between the spectral dimension of the three-dimensional effective QEG space-times with the one measured in Causal Dynamical Triangulations (CDT) \cite{Benedetti:2009ge}. Notably, the best-fit RG-trajectory reproduces the CDT data with approximately 1\% accuracy for the range of diffusion times where the simulation data is reliable. The comparison of $\cD_s^{\rm CDT}(T)$ with $\cD_s^{\rm QEG}(T)$ in Fig.\ \ref{p.fit2} furthermore establishes that the present Monte Carlo simulations neither probe the semi-classical plateau nor the scaling regime of the NGFP. This confirms the cautious remark in ref.\ \cite{Benedetti:2009ge} that present day Monte Carlo simulations are unable to probe physics well below the Planck length. This assessment 
 also resolves the apparent contradiction between the extrapolation result $\lim_{T \rightarrow 0} \cD_s^{\rm CDT}(T) \approx 2$ and the QEG prediction $\lim_{T \rightarrow 0} \cD_s^{\rm QEG}(T) = 3/2$: The fit function employed in analyzing the Monte Carlo data can not be reliable extrapolated to $T=0$ and misses essential structures.

The same conclusion also holds true in four dimensions. Comparing the profiles of $\cD_s^{\rm QEG}(T)$ shown in Fig.\ \ref{Fig.spec} with the fitting functions used in the CDT \cite{janCDT} or EDT \cite{laiho-coumbe} simulations shows that all the Monte Carlo data points obtained are positioned on the {\it infrared} side of the turning point of the RG-trajectories underlying the QEG effective space-times. They neither probe the semi-classical plateau or the scaling regime of the NGFP. Performing the extrapolation of $\lim_{T \rightarrow 0} \cD_s^{\rm CDT}(T)$ based on the leading corrections to the classical regime does not reliably identify the signature of a non-Gaussian fixed point in $\cD_s(T)$. Depending on where the data is cut off, one obtains different tangents to the first crossover, which lead to widely different extrapolations for the value $d_s = \cD_s(T)|_{T=0}$. We believe that this is actually at the heart of the apparent mismatch in the spectral dimension for infinitesimal random walks reported from the CDT and EDT computations. 

In order to test our conjecture that the non-classical $\cD_s$-values found in the simulations are due to the ``low energy fractality'' predicted by QEG in the Einstein-Hilbert approximation
one could perform the following numerical experiment. One couples gravity to a matter field whose parameters are averaged such that the resulting $8\pi G_k T^k_{\m\n}$ is approximately $k$-independent\footnote{Of course, for this purpose one could also give a non-NGFP scale dependence to $T^k_{\m\n}$ and try to observe its impact on $\langle g_{\m\n} \rangle_k$. (Note that $G_k$ will be approximately constant for $k \lesssim m_{\rm Pl}$).} and much bigger than $\lb_k \langle g_{\m\n} \rangle_k$. For this case we expect that the effective geometry has no significant $k$-dependence, so that the fractal features disappear, and the spectral dimension equals the classical one. On the other hand, the dimensional reduction implied by the NGFP cannot be destroyed in this way. Assuming the fixed point is also present in the gravity-matter system, it enforces a scale-dependence upon $8 \pi G_k T^k_{\m\n}$ and the other terms in the effective field equation, which is precisely such that the solutions behave as $\langle g_{\m\n} \rangle_k \propto 1/k^2$, implying the dimensional reduction $\cD_s = d \rightarrow d/2$.

We close our discussion with the following remark. In \cite{Horavafit} the data set displayed in Fig.\ \ref{Fig.CDT} has been fitted to an anisotropic gravity model of Ho\v{r}ava-Lifshitz type. Comparing the quality of this fit with the residuals displayed in Fig.\ \ref{p.fit1}, we observe that both theories fit the CDT data with approximately equal quality. This is in particular remarkable, if one takes into account that approximating $\cD_s(T)$ within the flat-space Einstein-Hilbert truncation gives rise to two fit parameters only, instead of the three parameters of the anisotropic model. Thus it may be premature to conclude that the spectral dimension obtained from the triangulations unequivocally identifies the underlying continuum theory as an anisotropic gravity model. In this light it seems mandatory to improve the simulation data in order to pinpoint the pertinent fractal structures of space-time with sufficient accuracy to reliably identify the underlying continuum theory. 

\section*{Acknowledgments}
%
We are indebted to D.\ Benedetti and J.\ Henson for sharing their CDT data with us and providing many helpful explanations. We are also grateful to J.\ Ambj{\o}rn, R.\ Loll and G.\ Calcagni for helpful discussions and S.\ Rechenberger for a careful reading of the manuscript. M.R.\ also thanks G.\ Dunne for inspiring conversations. The research of F.S.\ is supported by the Deutsche Forschungsgemeinschaft (DFG)
within the Emmy-Noether program (Grant SA/1975 1-1).

\begin{appendix}
\end{appendix}


\end{document}